\documentclass{ws-procs9x6}

\IfFileExists{srcltx.sty}{\usepackage[active]{srcltx}}

\begin{document}

\title{Dark matter's X-files}

\author{Alexander Kusenko}

\address{Department of Physics and Astronomy, University of California, Los
Angeles, CA 90095-1547, USA}


\begin{abstract}
Sterile neutrinos with keV masses can constitute all or part of the
cosmological dark matter.  The electroweak-singlet fermions, which are usually
introduced to explain the masses of active neutrinos, need not be heavier than 
the electroweak scale; if one of them has a keV-scale mass, it can be the
dark-matter particle, and it can also explain the observed pulsar kicks. The
relic sterile neutrinos could be produced by several different mechanisms. If
they originate primarily from the Higgs decays at temperatures of the order of
100 GeV, the resulting dark matter is much ``colder'' than the warm dark matter
produced in neutrino oscillations. The signature of this form of dark matter is
the spectral line from the two-body decay, which can be detected by the X-ray
telescopes.  The same X-rays can have other observable manifestations, in
particular, though their effects on the formation of the first stars. 

\end{abstract}


\section{Neutrino masses and the emergence of sterile neutrinos}

Most discoveries in particle physics amount to either a measurement of some
parameter related to a known particle, or a detection of some new degrees of
freedom,  new particles.  The discovery of the neutrino mass\cite{review} is
both.  Not only is it a measurement of the non-zero mass, but it also implies
the existence of some additional, SU(2) singlet fermions, ``right-handed''
neutrinos.  The corresponding particles can be made very heavy even for small
masses of the active neutrinos (the seesaw mechanism~\cite{seesaw}), but they
can also be light, in which case they are called sterile neutrinos.  The name
{\em sterile neutrino} was coined by Bruno~Pontecorvo, who
hypothesized the existence of the right-handed neutrinos in a seminal
paper~\cite{pontecorvo}, in which he also considered vacuum neutrino
oscillations in the laboratory and in astrophysics, the lepton number
violation, the neutrinoless double beta decay, some rare processes, such as
$\mu \rightarrow e \gamma$, and several other questions that have dominated the
neutrino physics for the next four decades.   Most models of the neutrino
masses introduce sterile (or right-handed) neutrinos to generate the masses of
the ordinary neutrinos via the seesaw mechanism~\cite{seesaw}.  The seesaw
lagrangian 
\begin{equation}
{\cal L}
  = {\cal L_{\rm SM}}+\bar N_{a} \left(i \gamma^\mu \partial_\mu 
\right )
N_{a}
  - y_{\alpha a} H \,  \bar L_\alpha N_{a} 
  - \frac{M_{a}}{2} \; \bar {N}_{a}^c N_{a} + h.c. \,,
\label{lagrangianM}
\end{equation}  
where ${\cal L_{\rm SM}}$ is the lagrangian of the Standard
Model, includes some number $n$  of singlet neutrinos $N_a$ ($a=1,...,n$) 
with Yukawa couplings $ y_{\alpha
a}$.  Here $H$ is the Higgs doublet and $L_\alpha$
($\alpha=e,\mu,\tau$) are the lepton doublets. Theoretical considerations do
not constrain the number $n$ of sterile neutrinos. In particular, there is no
constraint based on the anomaly cancellation because the sterile fermions do
not couple to the gauge fields. The experimental limits exist only for the
larger mixing angles~\cite{sterile_constraints}. To explain the neutrino masses
inferred from the atmospheric and solar neutrino experiments, $n=2$ singlets
are sufficient~\cite{2right-handed}, but a greater number is required if the
lagrangian (\ref{lagrangianM}) is to explain the r-process
nucleosynthesis~\cite{r}, the pulsar kicks~\cite{ks97,Kusenko:1998yy} and the
strength of the supernova explosion~\cite{Fryer,Hidaka:2006sg}, as well as 
dark matter~\cite{dw,Fuller,shi_fuller,smirnov,nuMSM}.  The same particle can
play an important role in the formation of the first stars~\cite{reion} and
other astrophysical phenomena~\cite{biermann_munyaneza}. 

The scale of the right-handed Majorana masses $M_{a}$ is unknown; it can be
much greater than the electroweak scale~\cite{seesaw}, or it may be as low as a
few eV~\cite{deGouvea:2005er}.  
The seesaw mechanism~\cite{seesaw} can explain the smallness of the neutrino
masses in the presence of the Yukawa couplings of order one if the 
Majorana masses $M_a$ are much larger than the electroweak scale. Indeed, in
this case the masses of the lightest neutrinos are suppressed by the ratios $
\langle H \rangle/M_a$. However, the origin of the Yukawa couplings remains
unknown, and there is no
experimental evidence to suggest that these couplings must be of order 1. In
fact, the Yukawa couplings of the charged leptons are much smaller than 1. For
example, the Yukawa coupling of the electron is as small as $(10^{-6})$.  

The Majorana mass can arise from the Higgs mechanism\cite{Chikashige:1980ht}.
For example, let us consider the following modification of the lagrangian
(\ref{lagrangianM}): 
\begin{equation} 
{\cal L}
   =   {\cal L}_{0}+\bar N_{a} \left(i \gamma^\mu \partial_\mu 
\right )
N_{a}  - y_{\alpha a} H \,  \bar L_\alpha N_{a}  - \frac{h_a}{2} \, S \,
\bar {N}_{a}^c N_{a} 
 +V(H,S) + h.c. \,, 
\label{lagrangianS}
\end{equation}
where $ {\cal L}_{0}$ includes the gauge and kinetic terms of the Standard
Model, $H$ is the Higgs doublet, $S$ is the real boson which is SU(2)-singlet,
$L_\alpha$ ($\alpha=e,\mu,\tau$) are the lepton doublets, and $ N_{a}$
($a=1,...,n$) are the additional singlet neutrinos.  
%
After the symmetry breaking, the Higgs doublet and singlet fields each develop
a VEV, $\langle H\rangle= v_0=247$~GeV, $\langle S\rangle= v_1$, and the
singlet neutrinos acquire the Majorana masses $ M_a = h_a v_1$.   As
discussed below, this model
is suitable for generating dark matter in the form of sterile neutrinos.

\section{Dark matter in the form of sterile neutrinos}

Sterile neutrino is a dark matter candidate. Since the singlet fermions are
introduced anyway to explain the observed neutrino masses, one can ask whether
the same particles can be the dark matter.  Because of the small Yukawa
couplings, the keV sterile neutrinos are out of equilibrium at high
temperatures.  However, there are several ways in which the relic population of
sterile neutrinos can be produced.  

\begin{itemize}
 \item Sterile neutrinos can be produced from neutrino oscillations as was
proposed by Dodelson and Widrow (DW)\cite{dw}.  If the lepton asymmetry is
negligible, this scenario appears to be in conflict with a
combination of the X-ray bounds~\cite{x-rays} and the Lyman-$\alpha$
bounds~\cite{viel,silk}, although it is possible to evade this constraint if
the lepton asymmetry of the universe is greater than
$(10^{-3})$~\cite{shi_fuller}.  On the other hand, observations of dwarf
spheroids point to a non-negligible free-streaming length for dark
matter\cite{wdm}, which favors warm dark matter.  It is also possible that the
sterile neutrinos make
up only a fraction of dark matter~\cite{silk}, in which case they can still be
responsible for the observed velocities of pulsars~\cite{ks97,Kusenko:2006rh}.

\item The bulk of sterile neutrinos could be produced from decays of $S$
bosons at temperatures above the $S$ boson mass, $T\sim
100$~GeV\cite{Kusenko:2006rh}.  In this case,  the Lyman-$\alpha$ bounds on the
sterile neutrino mass are considerably weaker than in the DW case because the
momenta of the sterile neutrinos are red-shifted as the universe cools down
from $T\sim 100$~GeV. 

\item Sterile neutrinos can be produced from their 
coupling to the inflaton\cite{shaposhnikov_tkachev}, or the
radion\cite{Kadota:2007mv}.  

\end{itemize}

It is important to note that only in the first case, the DW scenario,  the
dark matter abundance is directly related to the mixing angle.  In contrast,
if the relic population of sterile neutrinos arises from the Higgs decays,
their abundance is determined by the coupling $h$ in eq.~(\ref{lagrangianS}),
while the mixing angle is controlled by a different coupling $y$.  

We also note that both models, with largangians ~(\ref{lagrangianM})
and~(\ref{lagrangianS}), allow for some production of sterile neutrinos from
oscillations, but in the case of the singlet Higgs decays~(\ref{lagrangianS})
the bulk of the sterile dark matter could be produced at  $T\sim 100$~GeV, 
regardless of the value of the mixing angle, which can be vanishingly small. 

 Indeed, if the couplings of $S$ to $H$ are large enough, while $h<10^{-6}$,
the $S$ boson can be in equilibrium at temperatures above its mass, while the
sterile neutrino with a small mixing angle can be out of equilibrium at all
times.  Since $S$ is in thermal equilibrium at
high temperatures, some amount of sterile neutrinos can be produced in 
decays $S\rightarrow NN$: 
\begin{equation}
\Omega_{\nu_s} =  0.2 \left ( \frac{33}{\xi} \right )
\left ( \frac{h}{ 1.4 \times 10^{-8} } \right )^3
\left ( \frac{ \langle S \rangle }{\tilde{m}_{_S} } \right )
\label{Omega_w_VEVs}, 
\end{equation}
where $\xi $ is the change in the number density of sterile neutrinos
relative to $T^3$ due to the dilution taking place as the universe cools. For
example, in the Standard Model, the reduction in the number of effective
degrees of freedom that occurs during the cooling from the temperature $T\sim
100$~GeV to a temperature 1~keV causes the entropy increase and the
dilution of any species out of equilibrium by factor $\xi\approx 33$.  

At the same time, the sterile neutrino mass is determined by the VEV of $S$: 
\begin{equation}
h \langle S \rangle \sim {\rm keV} \ \ \ \Longrightarrow \ \ \ 
\langle S \rangle \sim \frac{\rm keV}{h} \sim 10^2 {\rm GeV}
\end{equation}
Based on the required values of $\Omega_s$ and the mass, we conclude that
the Higgs singlet should have a VEV at the electroweak scale. The dark matter
abundance, as in eq.~(\ref{Omega_w_VEVs}), was first computed for a model in
which the $S$ field served as the inflaton with a potential adjusted to have
$\langle S \rangle \gg m_S$, and much smaller values of $h, \xi$ were
considered\cite{shaposhnikov_tkachev}.  The alternative
possibility\cite{Kusenko:2006rh}, $\langle S \rangle \sim m_S$, which some may
find more natural, places the singlet Higgs right at the elecroweak scale,
which has important implications for the LHC\cite{singlet_higgs_LHC}.

\section{X-ray detection of relic sterile neutrinos}

\begin{figure}[ht!]
  \centering
  \includegraphics[width=10cm]{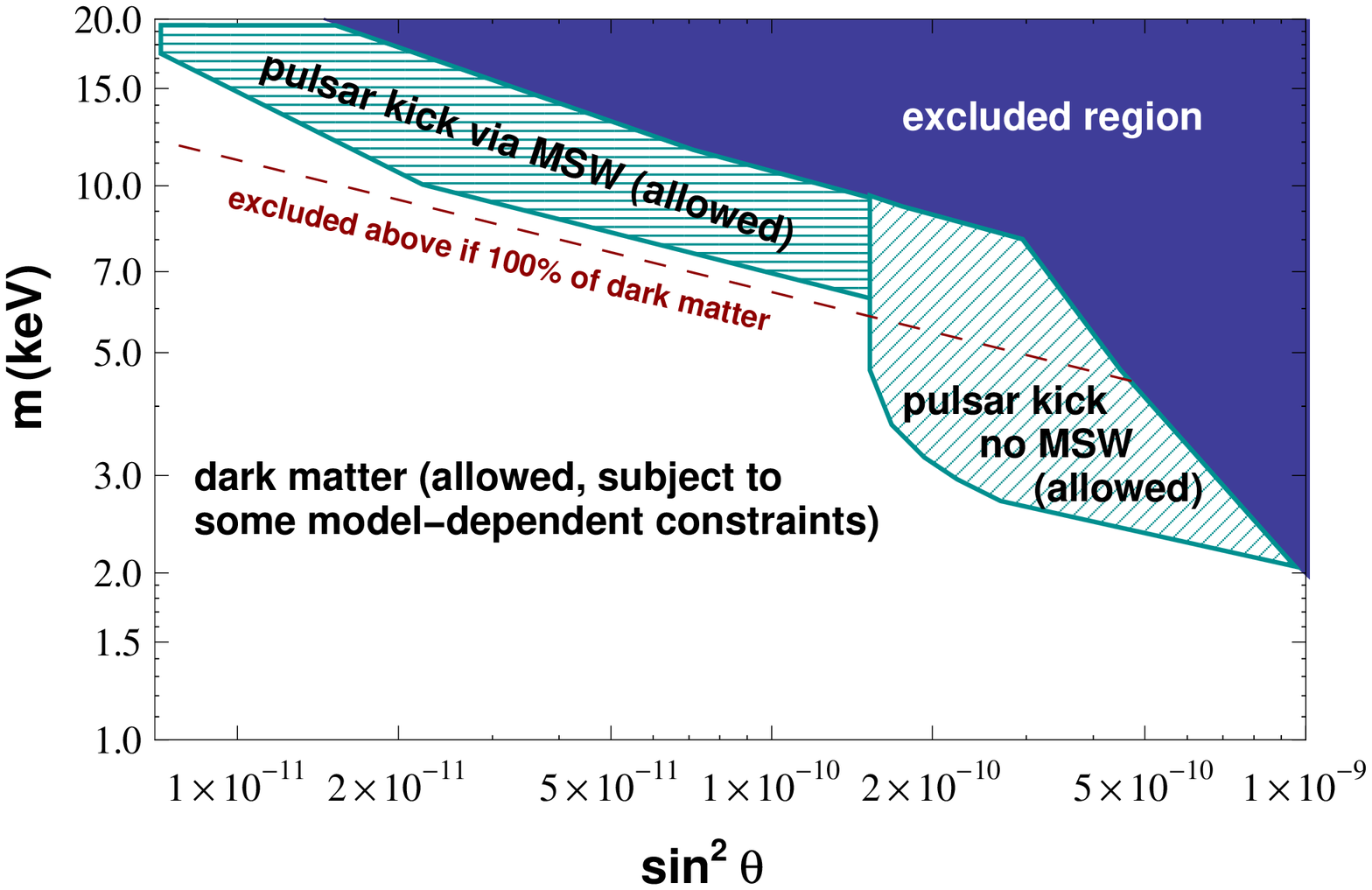}
  \caption{The solid excluded region is based on a combination of the X-ray 
and the small-scale structure bounds~\cite{silk}; it applies even if
sterile neutrinos constitute only a fraction of dark matter. The dashed line
shows the X-ray bound under the assumption that sterile neutrinos make up all
the dark matter.   Additional bounds from structure formation may apply,
depending on the free-streaming length, whose relation with the particle mass
depends on the production scenario. The region for the pulsar kicks shown here
is based on the re-analyzes of the earlier results\cite{ks97}, which will be
reported in an upcoming paper.}
  \label{fig:range}
\end{figure}

The relic sterile neutrinos can decay into the lighter neutrinos and an the
X-ray photons~\cite{pal_wolf}, which can be detected by the X-ray
telescopes~\cite{x-rays}. The X-ray flux depends on the sterile neutrino
abundance.  If all the dark matter is made up of sterile neutrinos
($\Omega_s\approx 0.2 $), then the limit on the mass and the mixing angle is
given by the dashed line in Fig.~\ref{fig:range}. 
However, the interactions in the lagrangian (\ref{lagrangianM}) cannot produce
such an $ \Omega_s= 0.2 $ population of the sterile neutrinos for the masses
and mixing angles along this dashed line, unless the universe has a relatively
large lepton asymmetry~\cite{shi_fuller}.  If the lepton asymmetry is small,
the interactions in eq.~(\ref{lagrangianM}) can produce the relic sterile
neutrinos via the neutrino oscillations off-resonance at some sub-GeV
temperature~\cite{dw}. This mechanism provides the lowest possible abundance
(except for the low-temperature cosmologies, in which the universe is never
reheated above a few MeV after inflation~\cite{low-reheat}).  The
model-independent bound~\cite{Kusenko:2006rh,silk} based on this scenario is
shown as a solid (purple) region in Fig.~\ref{fig:range}.  It is based on the
flux limit from X-ray observations\cite{x-rays} and the state-of-the-art
calculation of the sterile neutrino production by
oscillations\cite{Asaka:2006rw}. 

\section{X-rays and the formation of the first stars}

 The X-ray photons from sterile neutrino decays in the early universe could
have affected the star formation.  Although these X-rays alone are not
sufficient to reionize the universe, they can catalyze 
the production of molecular hydrogen and speed up the star
formation~\cite{reion}, which, in turn, would cause the reionization.
Molecular hydrogen is a very important
cooling agent, necessary for the collapse of primordial gas clouds that gave
birth to the first stars.  The fraction of molecular hydrogen must exceed a
certain minimal value for the star formation to begin~\cite{Tegmark:1996yt}. 
The reaction H+H$\rightarrow$H$_2 +\gamma$ is very slow in comparison with the
combination of reactions 
\begin{eqnarray}
{\rm H}^{+}+{\rm H}  & \rightarrow & {\rm H}_2^++ \gamma , \\  
{\rm H}_2^{+}+{\rm H} & \rightarrow & {\rm H}_2+{\rm H}^+, 
\end{eqnarray} %
which become possible if the hydrogen is ionized.  Therefore, the ionization
fraction determines the rate of molecular hydrogen production.  If dark
matter is made up of sterile neutrinos, their decays produce a sufficient flux
of photons to increase the ionization fraction by as much as two orders of
magnitude~\cite{reion}.  This has a dramatic effect on the
production of molecular hydrogen and the subsequent star formation.

Decays of the relic sterile neutrinos during the dark ages could produce an
observable signature in the 21-cm background~\cite{Valdes:2007cu}. It can be
detected and studied by such instruments as the Low Frequency Array (LOFAR),
the 21 Centimeter Array (21CMA), the Mileura Wide-field Array (MWA) and the
Square Kilometer Array~(SKA).

\subsection{Sterile neutrinos and the supernova}

Sterile neutrinos with masses below several MeV can be produced in the
supernova explosion; they can play an important role in the
nucleosynthesis\cite{r}, as well as in generating the supernova asymmetries and
the pulsar kicks\cite{ks97}.   Since the sterile
neutrinos interact with nuclear matter very weakly, they can be very efficient
at transporting the heat in the cooling proto-neutron star, altering the
dynamics of the supernova~\cite{Hidaka:2006sg}.  This could lead to an
enhancement of the supernova explosion. An additional enhancement can come
from the increase in convection in front of the neutron star propelled by the
asymmetric emission of sterile neutrinos~\cite{Fryer}. 

\subsubsection{The pulsar kicks}

The observations of
neutrinos from SN1987A constrain the amount of energy that the sterile
neutrinos can take out of the supernova, but they are still consistent with the
sterile neutrinos that carry away as much as a half of the total energy of the
supernova.  A more detailed analysis shows that the emission of sterile
neutrinos from a cooling newly born neutron star is anisotropic due to the
star's magnetic field~\cite{ks97}.  The anisotropy of this emission can
result in a recoil velocity of the neutron star as high as $\sim 10^3$km/s.
While both the active and the sterile neutrinos are produced with some
anisotropy, the asymmetry in the amplitudes of active neutrinos is quickly
washed out in multiple scatterings as these neutrinos diffuse out of the star
in the approximate thermal equilibrium~\cite{Kusenko:1998yy}.  In contrast, the
sterile neutrinos are emitted from the supernova with the asymmetry equal to
their production asymmetry.  Hence, they give the recoiling neutron star a
momentum, large enough to explain the pulsar kicks for the
neutrino emission anisotropy as small as a few per cent~\cite{ks97}.  
This mechanism can be the explanation of the observed pulsar 
velocities. The range of masses and mixing angles
required to explain the pulsar kicks is shown in Fig.~\ref{fig:range}.  

The pulsar kick mechanism based on the sterile neutrino emission has several
additional predictions~\cite{ks97}: 
\begin{itemize}
 \item the kick velocities were predicted to correlate with the axis of
rotation\cite{ks97}; recently, this spin-kick correlation was 
confirmed by the observations\cite{spin-kick_correlation}
 \item the kick should last 10 to 15 seconds, while the protoneutron star is 
cooling by the emission of neutrinos, but the onset of the kick can be delayed
by a few seconds, depending on the mass and mixing angles~\cite{ks97}; this
delayed kick can be tested using the observational data\cite{sp} 
\item neutrino-driven kicks can deposit additional energy behind the
supernova shock~\cite{Fryer,Hidaka:2006sg}, and they are expected to produce
asymmetric jets with the stronger jet pointing \textit{in the same direction}
as the neutron star velocity~\cite{Fryer}. 
\end{itemize}

\section{Conclusions}

The fundamental physics responsible for the neutrino masses is likely to
involve some additional SU(2)-singlet fermions, or sterile neutrinos.  The
Majorana masses of these states can range from a few eV to some values well
above the electroweak scale.  A sterile neutrino with a keV mass is a viable
dark matter candidate.  One can discover the relic sterile neutrinos
using the X-ray observations.   The same neutrinos can be produced in the
supernova explosions, and the anisotropy in their emission can explain the
observed pulsar velocities.    The X-rays from the sterile neutrino decays can
play an important role in the production of molecular hydrogen, which is
necessary for the formation of the first stars. 

This work was supported in part by the DOE grant DE-FG03-91ER40662 and by the
NASA ATP grants NAG~5-10842 and NAG~5-13399.  The author appreciates the
hospitality of Aspen Center for Physics, where part of this work was done.

\end{document}